\documentclass[useAMS]{mn2e}

\usepackage{graphicx}
 \usepackage{amssymb}
 \usepackage{url}
\usepackage{times}
\def\spose#1{\hbox to 0pt{#1\hss}}
\newcommand\lsim{\mathrel{\spose{\lower 3pt\hbox{$\mathchar"218$}}
     \raise 2.0pt\hbox{$\mathchar"13C$}}}
\newcommand\gsim{\mathrel{\spose{\lower 3pt\hbox{$\mathchar"218$}}
     \raise 2.0pt\hbox{$\mathchar"13E$}}}
\def\ltsima{$\; \buildrel < \over \sim \;$}
\def\lsim{\lower.5ex\hbox{\ltsima}}
\def\gtsima{$\; \buildrel > \over \sim \;$}
\def\gsim{\lower.5ex\hbox{\gtsima}}

\def\ergs{{\rm\thinspace erg \thinspace s^{-1}}}
\def\kms{{\rm\thinspace km \thinspace s^{-1}}}

\usepackage[normalem]{ulem}

\title[Disc--torus connection]
{The WISE view of the disc--torus connection in $z\sim$0.6 Active Galactic Nuclei}
\author[Calderone, G.]
{G. Calderone$^{1,3}$\thanks{E--mail:
giorgio.calderone@mib.infn.it}, T. Sbarrato$^{2,3}$, G. Ghisellini$^3$ \\
$^1$ Univ. di Milano Bicocca, Dip. di Fisica G. Occhialini, Piazza della Scienza 3, I--20126 Milano, Italy \\
$^2$ Univ. dell'Insubria, Dipartimento di Fisica e Matematica, Via Valleggio 11, I--22100 Como, Italy \\
$^3$ INAF -- Osservatorio Astronomico di Brera, via E. Bianchi 46, I--23807 Merate, Italy \\
}

\begin{document}

\pagerange{\pageref{firstpage}--\pageref{lastpage}} \pubyear{2012}

\maketitle
\label{firstpage}

\begin{abstract}
We selected all radio--quiet AGN in the latest release of the Sloan
digital sky survey quasar catalog, with redshift in the range
0.56--0.73. About 4000 ($\sim$80\%) of these have been detected in all
four IR--bands of WISE (Wide--field Infrared Survey Explorer).  This
is the largest sample suitable to study the disc--torus connection.
We find that the torus reprocesses on average $\sim$1/3--1/2 of the
accretion disc luminosity.
\end{abstract}



\section{Introduction}
\label{sec-intro}

Since the observations of NGC 1068 in polarized light by Antonucci \&
Miller (1985), showing the presence of broad permitted lines in
emission, the idea of the unification scheme of radio--quiet Seyfert
galaxies and quasars emerged (for reviews see Antonucci 1993; Robson
1996; Peterson 1997; Wills 1999; Krolik 1999).  The simplest version
of the scheme assumes the presence of a dusty ``torus'' surrounding
the central regions of the Active Galactic Nucleus (AGN) intercepting
a fraction of the illuminating accretion disc radiation and
re--emitting it in the infrared.  If the absorption is due to dust,
there is a natural temperature scale in the system, since dust
sublimates for temperatures greater than $\sim 1500$ K, corresponding
to a peak in the corresponding black body spectrum at $\nu_{\rm p} =
3.93\ kT/h\sim 1.2 \times 10^{14}$ Hz (or $\lambda_{\rm p}\sim 2 \mu$m;
the 3.93 factor is appropriate for the peak in the $\nu L_\nu$
spectrum).  The torus origin, stability, structure (see e.g. Krolik \&
Begelman 1988) and its very presence in both highly luminous
radio--quiet quasars and in low luminosity radio loud sources is under
debate.  The amount of reprocessed IR radiation seems to become
smaller for larger optical luminosity in radio--quiet objects
(i.e. ``receding torus'', Lawrence 1991), while, for radio sources,
the absence of broad emission lines in low power FR I radio--galaxies
(and BL Lacs) could be intrinsic, and not due to an obscuring torus
(Chiaberge, Capetti \& Celotti 1999).  Along the years, the idea of a
simple and uniform ``doughnut'' around the accretion disc has been
replaced by a clumped material, possibly outflowing (or inflowing), as
envisaged and modeled by many authors (see e.g. Elvis 2000; Risaliti,
Elvis \& Nicastro 2002; Elitzur \& Shlosman 2006; Nenkova 2008).

The existence of the unifying scenario based upon intrinsically equal
but observationally different AGN is also at the base of synthesis
models of the X--ray background (Setti \& Woltjer 1989; Madau,
Ghisellini \& Fabian 1994; Comastri et al. 1995; Gilli, Comastri \&
Hasinger 2007), since also the X--rays are partly absorbed, and partly
(Compton) reflected by the torus (Ghisellini, Haardt \& Matt 1994):
for large viewing angles, the observed X--ray emission becomes very
hard, as required to fit the X--ray background.

The covering factor of the absorbing material forming the ``torus'' is
not well known.  Estimates come from direct observations of optical
and IR AGN, as well from statistical considerations concerning the
number of type 1 and type 2 AGN.  In the first case, the studies were
hampered up to now by the relatively small samples of objects
(especially in the IR) suitable for a combined study (see e.g. Landt
et al. 2011 for a sample of 23 objects observed spectroscopically in
the optical and in the IR, down to $\sim 3\mu$m).

In order to study the accretion disc--torus connection in AGN, we need
to collect the largest group of radio--quiet AGN with reliable
detections of the IR luminosity and an optical spectrum to
characterize the accretion disc features.  The Sloan Digital Sky
Survey (SDSS; York et al.\ 2000) and the Wide--field Infrared Survey
Explorer (WISE; Wright et al.\ 2010) are the catalogs with the widest
number of objects in these two bands, hence they are the most
appropriate for our study.  WISE provided photometric observations in
4 IR bands (3.4, 4.6, 12 and 22 $\mu$m) for half a billion sources
(all sky) with fluxes larger than 0.08, 0.11, 1 and 6 mJy in
unconfused regions on the ecliptic in the four bands.  The sensitivity
improves toward the ecliptic poles due to denser coverage and lower
zodiacal background\footnote{Cutri et al. 2012:
  \url{http://wise2.ipac.caltech.edu/docs/release/allsky} }.

We adopt a flat cosmology with $H_0=71$ km s$^{-1}$ Mpc$^{-1}$ and
$\Omega_{\rm M}=0.27$.

\section{Sample selection}
\label{sec-sample}

We consider the fifth edition of the SDSS Quasar Catalog (Schneider et
al. 2010), containing 105,783 quasars with magnitude smaller than
$M_{i' \, \rm band} = -22$ (i.e. $\nu L_\nu (5100$\AA$) \sim 10^{44}
\ergs$), at least an emission line with FWHM$>1000\kms$ and a reliable
spectroscopical redshift.  Continuum and line luminosities (as well as
many other spectral properties) in SDSS spectra have been measured by
Shen et al. (2011; hereafter S11).  We will use these data to estimate
optical bolometric luminosity of the sources, following two
independent methods. The first one relies on the 3000\AA\ continuum
luminosity: $L_{\rm bol}^{\rm iso} = 5.16 \times \nu L_\nu(3000$\AA)
(e.g. Elvis et al. 1994; Richards et al. 2006).  The second one will
be used for a consistency check of our results, and relies on H$\beta$
and MgII line luminosities (\S \ref{sec-lines}).  In the following we
will assume that the bolometric luminosity equals the accretion disc
luminosity.  The superscript ``iso'' reminds that they are derived
under the assumption of isotropic emission.

The requirement that all sources are observed in the rest frame
2500--5500 \AA\ range (to comprise both the H$\beta$ and MgII lines
and the continuum at 3000 \AA) sets our first selection criterion.
Given the wavelength coverage of the SDSS, we require a corresponding
redshift range: $0.56<z<0.73$.  The S11 catalog has also been
cross--correlated with the Faint Images of the Radio Sky at
Twenty--centimeter survey (FIRST; Becker et al. 1995) and hence S11
include in their sample the radio fluxes.  The flux limit of the FIRST
sample is $\sim$1 mJy at 1.4 GHz.  Therefore, we can select the
radio--quiet quasars as those objects observed by the FIRST without a
detectable radio flux.  The radio--quiet requirement ensures the
absence of a contamination from the jet in the wavelength intervals of
interest.  After the radio--quietness and the redshift selections, we
are left with 5122 sources.  We have cross--correlated this sample
with the WISE All--Sky source catalog requiring that the optical and
IR positions are closer than 2 arcsec (5082 sources), and selecting
only those objects with detections in all the four WISE IR--bands, to
have the most complete IR luminosity information.  This last selection
leaves us with a sample of 3965 WISE--detected, radio--quiet type 1
AGN in a redshift range $z$=0.56--0.73.

\section{Data analysis and results}
\label{sec-analysis}

For all the 3965 sources in our sample we computed the IR flux in the
four WISE bands by first transforming the observed (Vega) magnitudes
in the AB systems setting $m_{\rm AB} = m + \Delta m$, with $\Delta m$
given in Tab. \ref{tab-wiseflux}.  In the AB system, the
flux--magnitude relation is simply:
\begin{equation}
\label{eq-wiseflux}
\log F  \, =\,  - \, {m_{\rm AB}+48.6 \over 2.5}
\end{equation}
where the flux density is measured in erg s$^{-1}$  cm$^{-2}$ Hz$^{-1}$.
\begin{table}
  \begin{center}
    \caption{ Center wavelengths and frequencies of the four WISE
      bands, and value of $\Delta m$ needed to transform the
      magnitudes given in the Vega system to the AB one.  See Cutri et
      al. (2012) in {\tt
        http://wise2.ipac.scaltech.edu/docs/release/allsky} .
    }
    \label{tab-wiseflux}
    \begin{tabular}{cccc}
      \hline
      \hline
      band & $\lambda$ [$\mu$m] & log Freq [Hz] & $\Delta m$\\
      \hline
      1  &  3.435 & 13.35   & 2.699\\
      2  &  4.6   & 13.63   & 3.339\\
      3  & 11.56  & 14.03   & 5.174\\
      4  & 22.08  & 14.16   & 6.620 \\
      \hline\hline
    \end{tabular}
  \end{center}
\end{table}
The integrated IR luminosity is computed by assuming a power law
spectrum between two contiguous bands, and summing the contributions
in all the three intervals.  The slopes of the power laws are given
by:
\begin{equation}
\label{eq-slope}
\alpha_{\rm i+1,i} = { m_{\rm AB,i} - m_{\rm AB,i+1} \over 
2.5 \log (\lambda_{\rm i+1}/\lambda_{\rm i}) }
\end{equation}
The integrated luminosity in each interval is:
\begin{equation}
\label{eq-l}
L_{\rm i+1, i} = { \nu_{\rm i} L_{\nu_{\rm i}} 
\over 1-\alpha_{\rm i,i+1} }\, \left[ 1-(\nu_{\rm i+1}/\nu_{\rm i})^{1-\alpha_{\rm i, i+1}} \right]
\end{equation}
\begin{table}
  \begin{center}
    \caption{Mean and standard deviation of IR luminosities and
      spectral slopes in the four WISE bands for the whole sample, and
      the three subsample described in \S
      \ref{sec-analysis}. Luminosities are in units of erg
      s$^{-1}$.}
    \label{tab-reswise}
    \begin{tabular}{llll}
      \hline
      \hline
      &  Band  & log$\, \nu L_\nu$ & $\alpha$ \\
    \hline
    Whole  & 1 &   44.87$\pm$0.26  &      1.3$\pm$0.5\\
    sample & 2 &   44.92$\pm$0.29  &      0.9$\pm$0.3\\
           & 3 &   44.87$\pm$0.29  &      1.4$\pm$0.5\\
           & 4 &   44.99$\pm$0.27  &      --         \\
    \hline
    Sub A  & 1 &   44.88$\pm$0.20  &      1.2$\pm$0.4\\
           & 2 &   44.74$\pm$0.17  &      0.9$\pm$0.3\\
           & 3 &   44.77$\pm$0.15  &      1.5$\pm$0.4\\
           & 4 &   44.74$\pm$0.15  &      --         \\
    \hline
    Sub B  & 1 &   45.01$\pm$0.20  &      1.4$\pm$0.4\\
           & 2 &   44.91$\pm$0.19  &      0.9$\pm$0.2\\
           & 3 &   44.96$\pm$0.16  &      1.4$\pm$0.4\\
           & 4 &   44.91$\pm$0.15  &      --         \\
    \hline
    Sub C  & 1 &   45.15$\pm$0.22  &      1.6$\pm$0.3\\
           & 2 &   45.09$\pm$0.17  &      0.8$\pm$0.2\\
           & 3 &   45.16$\pm$0.15  &      1.2$\pm$0.4\\
           & 4 &   45.09$\pm$0.13  &      --         \\
    \hline\hline
    \end{tabular}
  \end{center}
\end{table}
Finally, the integrated luminosity is $L_{\rm IR}^{\rm iso} = L_{2,1}+
L_{3,2} + L_{4,3}$.  As discussed in \S \ref{sec-sample}, the
bolometric luminosity is computed as $L_{\rm bol}^{\rm iso} = 5.16
\times \nu L_\nu(3000$\AA).  Again, the ``iso'' superscript reminds
that these quantities are computed assuming isotropic emission.

\begin{figure}
\vskip -0.3cm
\includegraphics[width=9.05cm]{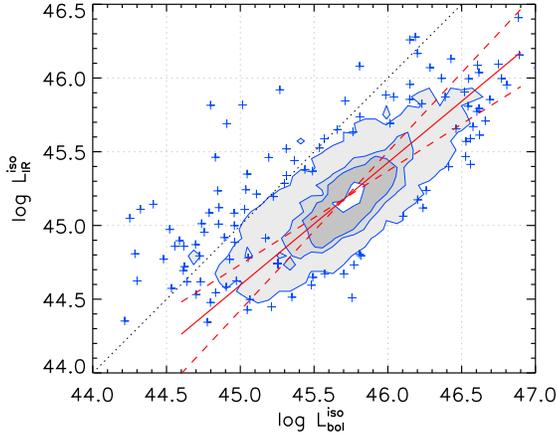} 
\vskip -0.3cm
\caption{Comparison of bolometric luminosity $L_{\rm bol}^{\rm iso}$
  and integrated IR luminosity $L_{\rm IR}^{\rm iso}$ as measured by
  WISE. Contour levels are at 10\%, 50\%, 68\% and 95\% of total
  source number (3965). The dotted line identifies equal
  luminosity. The solid line ($\log L_{\rm IR}^{\rm iso} \propto
  0.83\log L_{\rm bol}^{\rm iso}$) is the bisector of the two least
  squares fitting (dashed) lines.}
\label{fig-whole}
\end{figure}

The ratio $R=L_{\rm IR}^{\rm iso}/L_{\rm bol}^{\rm iso}$ is
approximately constant ($\sim 0.3$, Tab. \ref{tab-rescov},
Fig. \ref{fig-histo-covering}) and will be used in \S
\ref{sec-covfactor} to estimate the torus covering factor.  The
bolometric and IR luminosities of all sources show a well defined
correlation over at least 1.5 dex, as shown in Fig. \ref{fig-whole}.
We performed two least squares fits by taking at first $x = \log
L_{\rm bol}^{\rm iso}$ and $y = \log L_{\rm IR}^{\rm iso}$, then
inverting the variables.  We took the bisector as the best description
of the correlation: $\log L_{\rm IR}^{\rm iso} \propto 0.83\log L_{\rm
  bol}^{\rm iso}$.  The slope, being smaller than one, suggests that
IR luminosities become smaller at larger optical luminosity (receding
torus).  Similar results have been found using independent methods by
e.g. Arshakian (2005) and Simpson (2005).

To provide a deeper insight on the disc--torus connection we select
three subsamples according to $L_{\rm bol}^{\rm iso}$; we will refer
to these subsamples with letters A, B, C.  Tab. \ref{tab-reswise}
lists the $\nu L_\nu$ IR luminosities in the four WISE bands for the
whole sample and for the subsamples A, B, C, together with the average
spectral indices.  Tab. \ref{tab-rescov} reports the average and the
standard deviation of $L_{\rm bol}^{\rm iso}$ and $L_{\rm IR}^{\rm
  iso}$ together with their ratio $R=L_{\rm IR}^{\rm iso}/L_{\rm
  bol}^{\rm iso}$ for the whole sample and for the A, B, C subsamples.
For the latter, instead of the standard deviation of $L_{\rm bol}^{\rm
  iso}$, we give the logarithmic width of the considered luminosity
bin.  Note that sources in these subsamples account for only $\sim$1/3
of the entire sample.  Dropping 2/3 of the sample was necessary to
significantly separate the bolometric luminosity classes.

\begin{table*}
  \begin{center}
    \caption{Results of our analysis. Columns are: (1) sample; (2)
      number of sources in the sample; (3) mean bolometric luminosity
      in the sample; (4) width of luminosity bin ($^a$: value in the
      first row is the standard deviation); (5) mean and standard
      deviation of $L_{\rm IR}^{\rm iso}$ in the sample; (6) mean and
      standard deviation of parameter $R = L_{\rm IR}^{\rm iso} /
      L_{\rm bol}^{\rm iso}$; (7) range of covering factor
      (Eq. \ref{eq-c-r}); (8) range of torus opening angles; (9) range
      of Type 2 to Type 1 AGN count ratio (\#2/\#1). All means and
      standard deviations are computed using logarithmic values.}
    \label{tab-rescov}
    \begin{tabular}{lllllllll}
      \hline
      \hline
      Sample &  N src.  &  log $L_{\rm bol}^{\rm iso}$  & log $\Delta L_{\rm bol}^{\rm iso}$  & 
      log  $L_{\rm IR}^{\rm iso}$ & $R$  &  Cov. factor   &   $\theta_{\rm T}$
      &  $\#2/\#1$\\
      \hline
      \hline
      Whole &    3965  &    45.72 &   0.33$^a$  &    45.18$\pm$0.27  &     $0.29_{-0.11}^{+0.18}$  &   0.54--0.70   &   57--46  &  1.2--2.3\\
      \hline                                                                                  
      A     &     408  &    45.55 &   0.10      &    45.05$\pm$0.16  &     $0.31_{-0.10}^{+0.14}$  &   0.56--0.74   &   56--42  &  1.3--2.8\\
      B     &     569  &    45.80 &   0.10      &    45.22$\pm$0.16  &     $0.26_{-0.08}^{+0.12}$  &   0.51--0.66   &   59--49  &  1.0--1.9\\
      C     &     389  &    46.05 &   0.14      &    45.40$\pm$0.16  &     $0.22_{-0.07}^{+0.10}$  &   0.47--0.60   &   62--53  &  0.9--1.5\\
      \hline
      \hline
    \end{tabular}
  \end{center}
\end{table*}

\subsection{Consistency with broad emission lines}
\label{sec-lines}

\begin{figure}
\vskip -0.3cm
\includegraphics[width=9cm]{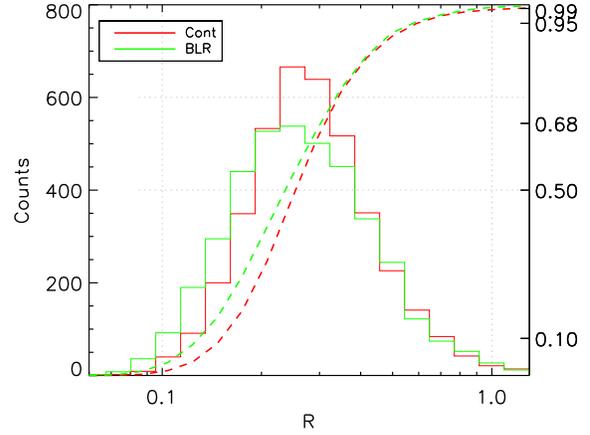} 
\vskip -0.3cm
\caption{Histogram of $R$, as computed using integrated IR luminosity
  $L_{\rm IR}^{\rm iso}$ and both the continuum--based (red solid
  line) and BLR--based (green solid line) bolometric
  luminosities. Dashed lines show the cumulative fraction (values on
  right axis).}
  \label{fig-histo-covering}
\end{figure}

The estimates given above do not take into account that the observed
optical continuum can include different components, besides the disc
emission.  As a consistency check, we use an alternative method to
derive the disc (bolometric) luminosity, by using the luminosities of
the H$\beta$ and the MgII broad lines, always present in the SDSS
spectra in our redshift selection.  For radiatively efficient discs,
indeed, the overall luminosity of the broad line region (BLR), $L_{\rm
  BLR}$, is a proxy of the disc luminosity $L_{\rm bol}$, since on
average $L_{\rm BLR} \sim \gamma L_{\rm bol}$, where the factor
$\gamma$ is directly connected to the BLR covering factor (see
e.g.\ Baldwin \& Netzer 1978; Smith et al.\ 1981).  In turn, $L_{\rm
  d}$ should be equal to $L_{\rm bol}$ (real, not isotropically
equivalent).  Estimates of $\gamma$ lie in quite large ranges,
historically between 0.002 and 0.35 (according to Baldwin \& Netzer),
but preferentially $<0.15$ (Smith et al.\ 1981).  Typically, an
average value $\gamma\sim0.05-0.1$ is assumed.  $L_{\rm BLR}$ can be
calculated from individual broad line luminosities, as in Celotti et
al.\ (1997).  Specifically, setting the Ly$\alpha$ flux contribution
to 100, the relative weights of the H$\alpha$, H$\beta$, MgII and CIV
broad lines are 77, 22, 34 and 63, respectively (Francis et
al.\ 1991).  The total broad line flux is fixed at 555.8.  Since all
our sources have measured H$\beta$ and MgII line luminosity, we
average for each object the estimates of $L_{\rm BLR}$:
\begin{equation}
L_{\rm BLR} = {1 \over 2} \left[
    \frac{555.8}{22} L({\rm H}\beta) + 
    \frac{555.8}{34} L({\rm MgII})
    \right]
\end{equation}
A good agreement between the continuum--based and the BLR--based
bolometric luminosities is obtained using $L^{\rm iso}_{\rm
  bol}=L_{\rm BLR}/0.041$.  This value is found for sources in the
highest luminosity subsample (C), for which we do not have spurious
contributions from components other than the disc (e.g. host galaxy).
Fig. \ref{fig-histo-covering} shows the histogram of the ratio $R$ for
all sources (green solid line).
%
%
This histogram not only has the same average of the distribution of
$R$ based on the 3000 \AA\ luminosity (which is expected, given our
assumptions), but the two distributions are similar for all $R$,
implying that broad lines are a good proxy to compute the total disc
luminosity, and that our estimates of $R$ are reliable.

\section{The covering factor of the torus}
\label{sec-covfactor}

Consider the simplest case of a doughnut--shaped torus with opening
angle $\theta_{\rm T}$, as measured from the symmetry axis.  The
covering factor $c$ is defined as:
\begin{equation}
c = \frac{\Omega_{\rm T}}{4 \pi} =  
{2\times 2\pi \int_{\theta_{\rm T}}^{\pi/2} \sin\theta d\theta \over 4\pi }
 = \cos \theta_{\rm T}
\end{equation}
We must relate $c$ to the observed ratio $R$, accounting for the
anisotropy of disc and torus emission.  Since the emission of
geometrically thin discs follows a $\cos\theta$ pattern, for a given
viewing angle $\theta_{\rm v}$ (calculated from the disc axis) the
ratio between the real disc luminosity $L_{\rm bol}$ and the isotropic
estimate $L_{\rm bol}^{\rm iso}$ is:
\begin{equation}
  \label{eq-anisotropy-disc}
{L_{\rm bol} \over L_{\rm bol}^{\rm iso}} =
{2\times 2\pi \int^{\pi/2}_0 \cos\theta \sin\theta d\theta \over 4\pi \cos\theta_{\rm v} }
= {1\over 2\cos\theta_{\rm v} }
\end{equation}
The ratio is smaller than unity for $\theta_{\rm v} < 60^\circ$, thus
for Type 1 AGN we likely have $L_{\rm bol} < L_{\rm bol}^{\rm iso}$.
We are not able to determine $\cos \theta_{\rm v}$ for each source,
but we can safely assert that $0 \le \theta_{\rm v} \le \theta_{\rm
  T}$, since we are dealing with Type 1 AGN.  Therefore a reasonable
estimate is:
\begin{equation}
  \cos\theta_{\rm v} \sim 
  \langle \cos\theta\rangle_{0-\theta_{\rm T}}
    =  {  \int_{0}^{\theta_{\rm T}} \cos\theta \sin\theta d\theta
      \over   \int_{0}^{\theta_{\rm T}} \sin\theta d\theta}
    = {1+\cos\theta_{\rm T} \over 2}
\end{equation}
A relation similar to Eq. \ref{eq-anisotropy-disc} for the torus
luminosity ($L_{\rm T} = L_{\rm IR}$) is currently unknown. However,
we can reasonably state that
\begin{equation}
  {L_{\rm bol} \over L_{\rm bol}^{\rm iso}} <
  {L_{\rm T} \over L_{\rm T}^{\rm iso}} < 1
\end{equation}
The lower limit corresponds to a thin disc--shaped emitting torus, the
upper limit to an isotropic emitting torus.  Both limits are rather
unrealistic: the torus is expected to show a lower degree of
anisotropy than the disc since we are able to detect radiation emitted
from the side (i.e. Type 2 AGN); also, the torus is hardly an
isotropic emitter since IR signatures are different in Type 1 and 2
AGN (Calderone et al., in prep.).  The above limits should then
bracket the real case.  The amount of disc radiation intercepted (and
re--processed) by the torus is:
\begin{equation}
{L_{\rm T} \over L_{\rm bol}}  = 
{\int^{\pi/2}_{\theta_{\rm T}} \cos\theta \sin\theta d\theta \over
\int^{\pi/2}_0 \cos\theta \sin\theta d\theta }
= \cos^2\theta_{\rm T} 
\end{equation}
Rearranging the previous equations, we find a relation between the
observable parameter $R = L_{\rm IR}^{\rm iso} / L_{\rm bol}^{\rm
  iso}$ and the covering factor $c$:
\begin{equation}
  \label{eq-c-r}
  \frac{c^2}{1 + c} < R < c^2
\end{equation}
This relation can be inverted to find the allowed range of $c$ and
$\theta_{\rm T}$, given a value of the observable parameter $R$.
Finally, the covering factor $c$ can be used to estimate the count
ratio between Type 1 and Type 2 AGN:
\begin{equation}
   c = \frac{\Omega_{\rm T}}{4 \pi} = \frac{\#2}{\#1 + \#2}
   \hspace{0.3cm} \Rightarrow \hspace{0.3cm}
   \frac{\#2}{\#1} = \frac{c}{1-c}
\end{equation}
The last three columns of Tab. \ref{tab-rescov} report the value of $c$,
$\theta_{\rm T}$ and \#2/\#1 corresponding to the observed values of
$R$ in all discussed samples.

\section{Discussion and conclusions}
\label{sec-discussion}

\begin{figure*}
  \vskip -0.3cm
  \includegraphics[width=18.5cm,clip,trim=22 12 30 23]{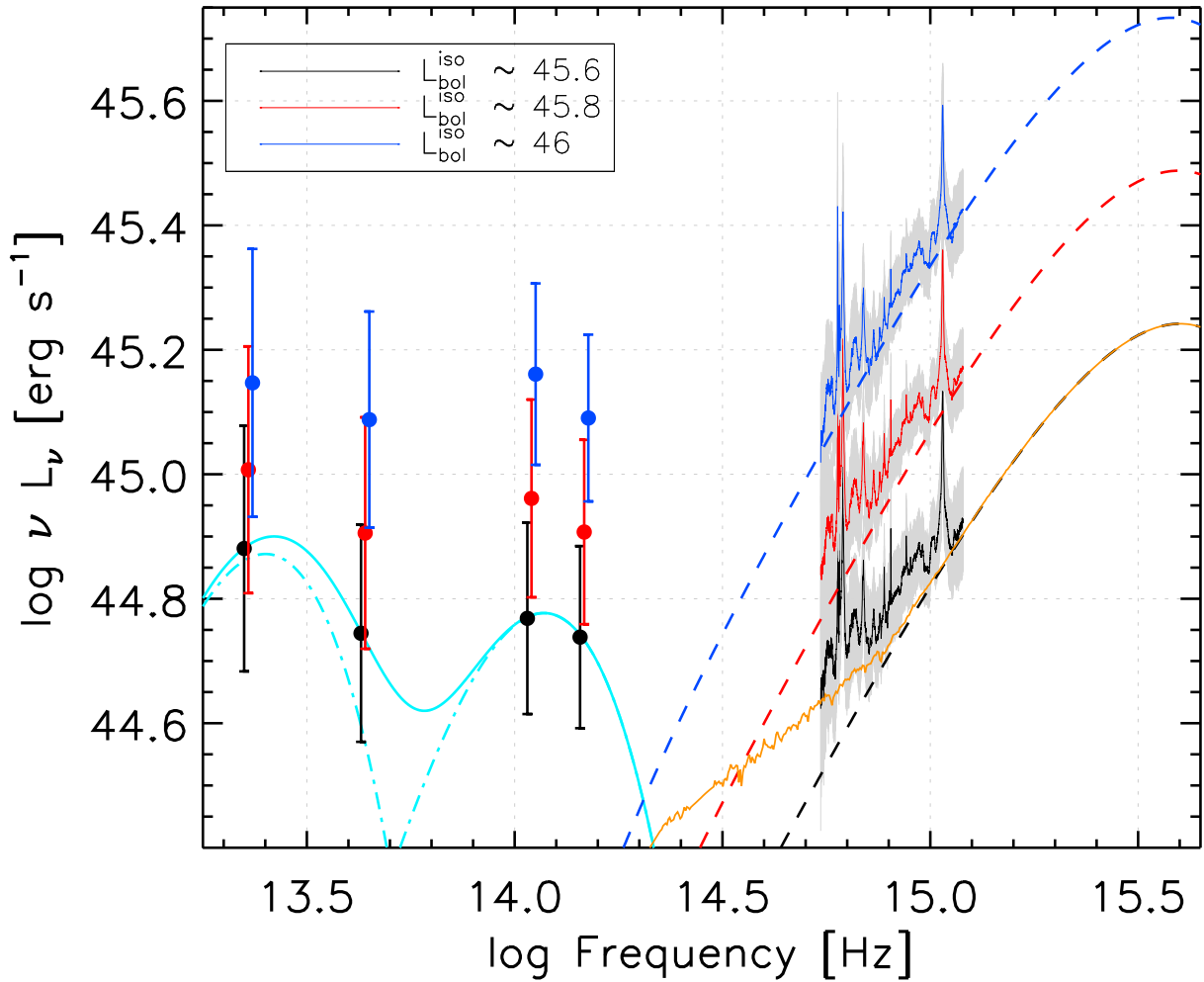}
  \caption{The disc--torus connection. AGN in our sample were divided
    in three subsamples according to their bolometric luminosity (see
    Tab. \ref{tab-rescov}), and associated to black, red and blue
    color respectively.  The logarithmic mean and standard deviation
    of IR luminosities in each subsample are computed using data from
    WISE, and displayed as filled circles and error bars (points are
    slightly displaced for a clearer view).  Composite optical/NUV
    spectra (solid color--coded lines) are computed as follows: SDSS
    spectra are transformed to rest--frame and de--reddened using
    Schlegel et al. (1998) and Pei (1992). The spectra are then
    rebinned to a common wavelength grid and a composite spectrum is
    computed as the geometric mean.  The gray shades indicate the 68\%
    level dispersion.  Standard Shakura \& Sunyaev (1973) accretion
    disc spectra fitting the composite spectra are shown with dashed
    lines.  The disc--torus connection is clearly visible in this
    figure, in which the torus luminosity in all four WISE bands
    follows the trend in accretion disc bolometric luminosity.
    Discrepancies between the composite spectrum in the lower
    luminosity subsamples (black and red) and the accretion disc
    spectrum may be due to the contribution of host galaxy starlight
    (Vanden Berk et al. 2011).  The yellow solid line shows the sum of
    the accretion disc spectrum (black dashed line) and the elliptical
    galaxy template from Mannucci et al. (2001) with a bolometric
    luminosity of log($L_{\rm host}$/erg s$^{-1}$) $\sim$ 44.3.  The
    IR points may be modeled as sum (solid line) of at least two black
    bodies with temperatures 308 K and 1440 K (dot--dashed lines), and
    luminosities log($L_{\rm torus,BB}$/erg s$^{-1}$) = 45.00 and
    44.91 respectively. }
  \label{fig-wise-sdss} 
\end{figure*}

The main result of our work is the determination of the average
covering factor of the torus using a very large data set.  The {\it
  observed} fraction of IR to bolometric, isotropically equivalent
optical luminosity is about 30\%.  This implies that the obscuring
torus covering factor is in the range 0.5--0.7 and that the opening
angle $\theta_{\rm T}$ is 40$^\circ$-- 60$^\circ$.  On average, our
sources emit in the IR a similar fraction of their bolometric
luminosities ($\sim 1/3$).  For each Type 1 AGN, there should be
between 1 and 3 Type 2 sources.  If there is a broad distribution of
covering factors (as suggested by Elitzur, 2012) our Type 1 sample may
be drawn preferentially from the lower end of the distribution.  In
this case our estimate of \#2/\#1 ratio is a lower limit.  The very
basic prediction of the unified model that the torus re--processes a
given amount of disc luminosity is verified (Fig. \ref{fig-whole} and
Fig. \ref{fig-histo-covering}). The dispersion of this fraction is
remarkably small, being at most a factor of 2. The broad--band
spectral energy distribution (SED) from IR to near--UV are expected to
be quite similar among Type 1 AGN.  A hint of the ``receding torus''
hypothesis is found in Fig. \ref{fig-whole}, with $\log L_{\rm
  IR}^{\rm iso} \propto 0.83\log L_{\rm bol}^{\rm iso}$.  In
Fig. \ref{fig-wise-sdss} we show both data and model for a
prototypical broad--band SED, in the three luminosity classes
considered above (coded with colors).  For each subsample we also
compute a composite spectrum using SDSS spectra.

At IR wavelengths the torus emission dominates.  Spectral indices
between the four WISE bands are very similar for different overall
luminosities (Tab. \ref{tab-reswise}).  Despite the rather poor
coverage, it appears that the IR emission is structured with at least
two broad bumps.  Such features are easily modeled by the
superposition of two black bodies with temperatures of $\sim$300 K and
$\sim$1500 K respectively.  A na\"\i ve interpretation is to consider the
hotter one as emitted from the hot part of the torus facing the disc,
at the dust sublimation temperature.  The colder one would come from
the cooler outer side of the torus.  This should be the region visible
also in Type 2 AGN.

The underlying optical continua are well described by a standard
Shakura \& Sunyaev (1973) accretion disc spectrum.  The dashed lines
in Fig. \ref{fig-wise-sdss} are the models of three accretion discs
having the same bolometric luminosity as the spectra in the subsample,
and masses $1.7 \times 10^8$, $2.3 \times 10^8$ and $3.4 \times 10^8$
M$_{\sun}$ respectively, grossly in agreement with the (virial) masses
calculated in S11.  The WISE data points (in $\nu L_\nu$) lie a factor
$\sim$3 below the disc peaks (at log($\nu$/Hz)$\sim$15.5).  This
factor corresponds to the value $\sim$1/3 found in
Tab. \ref{tab-rescov} and Fig. \ref{fig-histo-covering}.  The
composite spectra follow closely the accretion disc continuum in all
but the lowest luminosity subsample, in which some other component is
present at frequencies below log($\nu$/Hz)$<$14.9.  This further
component may be the starlight contribution from host galaxy (Vanden
Berk et al. 2011), as shown by the yellow line which is the sum of the
accretion disc spectrum and an appropriately scaled template for an
elliptical (quiescent) galaxy from Mannucci et al. (2001).  At higher
luminosity subsamples, the contribution from galaxy becomes relatively
less important.

\section*{Acknowledgements}
This publication makes use of data products from the Wide--field
Infrared Survey Explorer, which is a joint project of the University
of California, Los Angeles, and the Jet Propulsion
Laboratory/California Institute of Technology, funded by the National
Aeronautics and Space Administration.


\label{lastpage}

\begin{thebibliography}{99}

\bibitem[]{} Antonucci R. \& Miller J., 1985, ApJ, 297, 621

\bibitem[]{} Arshakian, T. G., 2005, A\&A, 436, 817A

\bibitem[]{} Baldwin J.A. \& Netzer H., 1978, ApJ, 226, 1

\bibitem[]{} Becker R.H., White R.L. \& Helfand D.J., 1995, ApJ, 450, 559

\bibitem[]{} Celotti A., Padovani P. \& Ghisellini G., 1997, MNRAS, 286, 415

\bibitem[]{} Chiaberge M., Capetti A \& Celotti A.,  1999, A\&A, 349, 77

\bibitem[]{} Comastri A., Setti G., Zamorani G. \& Hasinger G.,  1995, A\&A, 296, 1
	
\bibitem[]{} Elitzur M. \& Shlosman I., 2006, ApJ, 648, L101

\bibitem[]{} Elitzur M., 2012, ApJ, 747L, 33

\bibitem[]{} Elvis M., Wilkes B.J., McDowell J.C. et al., 1994, ApJS, 95, 1

\bibitem[]{} Elvis M., 2000, ApJ, 545, 63

\bibitem[]{} Francis P.J., Hewett P.C., Foltz C.B., Chaffee F.H., Weymann R.J. \& Morris S.L.,
		   1991, ApJ, 373, 465

\bibitem[]{} Ghisellini G., Haardt F. \& Matt G., 1994, MNRAS, 267, 743

\bibitem[]{} Gilli R., Comastri A. \& Hasinger G., 2007, A\&A, 463, 79

\bibitem[]{} Krolik J. \& Begelman M.C., 1988, ApJ, 392, 702

\bibitem[]{} Krolik J., 1999, {\it Active Galactive Nuclei}, Princeton: Princeton Univ. Press

\bibitem[]{} Landt H., Elvis M., Ward M., Bentz M.C., Korista K.T. \& Karovska M., 2011, MNRAS, 414, 218

\bibitem[]{} Lawrence A., 1991, MNRAS, 252, 586

\bibitem[]{} Madau P., Ghisellini G., Fabian A.C. 1994, MNRAS, 270, L17

\bibitem[]{} Mannucci F., Basile F., Poggianti B. M., Cimatti A.,
  Daddi E., Pozzetti L., Vanzi, L., 2001, MNRAS, 326, 745M

\bibitem[]{} Nenkova M., Sirocky M.M., Ivezic Z. \& Elitzur M., 2008, ApJ, 658, 147

\bibitem[]{} Pei Y.~C.,  1992, ApJ, 395, 130

\bibitem[]{} Peterson B.M., 1997, {\it Introduction to Active Galactic Nuclei}, Cambridge Univ. Press

\bibitem[]{} Richards G.T., Lacy M., Storrie--Lombardi L.J. et al., 2006, ApJS, 166, 470

\bibitem[]{} Risaliti G., Elvis M. \& Nicastro F., 2002, ApJ, 571, 234

\bibitem[]{} Robson I., 1996, {\it Active Galactic Nuclei}, John Wiley and Sons, Ltd. 
             in assoc. with Praxis Publishing, Ltd.

\bibitem[]{} Schlegel D.--J., Finkbeiner D.~P., Davis M.,  1998, ApJ, 500, 525

\bibitem[]{} Schneider D. P., Richards, G. T., Hall P. B. et al., 2010, AJ, 139, 2360S

\bibitem[]{} Setti G., \& Woltjer L., 1989, A\&A, 224, L1

\bibitem[]{} Shakura N.I. \& Sunjaev R.A., 1973, A\&A, 24, 337

\bibitem[]{} Shen Y., Richards G.T., Strauss M.A. et al.,  2011, ApJS, 194, 45, (S11)

\bibitem[]{} Simpson, C., 2005, MNRAS, 360, 565S

\bibitem[]{} Smith M.G., Carswell R.F., Whelan J.A.J et al., 1981, MNRAS, 195, 437

\bibitem[]{} Vanden Berk D. E. et al., 2001, AJ, 122, 549

\bibitem[]{} Wills B., 1999, {\it in Quasars and Cosmology}, 
             ASP Conf. Ser., vol. 162, p. 101, Ed. G. Ferland and J. Baldwin

\bibitem[]{} Wright E.L., Eisenhardt P.R.M., Mainzer A.K. et al., 2010, AJ, 140, 1868

\bibitem[]{} York D.G., Adelman J., Anderson J.E., Jr. et al., 2000, AJ, 120, 1579


\end{thebibliography}
\end{document}